# Open Area Path Finding to Improve Wheelchair Navigation

Anahid Basiri (ana.basiri@glasgow.ac.uk)

University of Glasgow, UK

This work was supported by the UK Research and Innovation. This work presented in this paper has been partially funded by the UK Research and Innovation (UKRI) Future Leaders Fellowship MR/S01795X/1

**ABSTRACT** Navigation is one of the most widely used mobile services many use on daily basis. However, as navigation services have been initially designed for in car use, they may require some customisations to accommodate the needs of other user groups, including pedestrians or wheelchair users. The movements of pedestrians and wheelchair users are not limited by the directed networks of streets and roads, but they may experience other limitations, including an increased impedance caused by the weather or pavement surface conditions, particularly for the latter group. This paper proposes and implements a novel path finding algorithm for open areas, i.e. areas with no networks of pathways, such as grasslands and parks where the conventional graph-based algorithms fail to calculate a practically traversable path. While the focus of the paper is developing a pathfinding algorithm for wheelchair users, the algorithm can be applied for other user groups, e.g. pedestrians. The proposed algorithm creates an enhanced visibility graph in the open area; accounts for obstacles and barriers; and calculates a suggested route based on the factors that are identified as important for wheelchair users. Factors, including slope, width, and surface condition of the routes, are recognised by mining the actual trajectories of wheelchairs users using trajectory mining and machine learning techniques. Unlike raster-based techniques, such as traditional grid or visibility graph-based algorithms, the proposed path finding algorithm allows the routing to be fully compatible with current transportation routing services and enables an efficient multimodal routing service. The implementations and evaluation of the service show at least a 76.4% similarity between the proposed algorithm outputs and actual wheelchair user trajectories and a higher users satisfaction level in comparison with routes suggested by widely-used services such as Google Maps.

## I. INTRODUCTION

While mobile navigation applications and routing services have become immensely popular, they have predominantly been designed and developed for in-vehicle use, which means they still need some customisation for other user groups. In the absence of alternatives, groups including pedestrians and wheelchair users will still use routing services designed for motor vehicles (as they are or with some minor changes), but this can cause problems as their different requirements and preferences are not fully catered for. For example, unlike drivers, pedestrians and wheelchair users, are not particularly limited to the road network and can move in any direction, go through buildings, or cross open areas such as parks and grasslands. This greater freedom of movement is not supported by most widely used navigation and routing services - which still calculate the path between an origin and destination using graph-based path finding algorithms.

**Figure 1.** The suggested path by OpenStreetMap for pedestrians between two points both located in a grassland

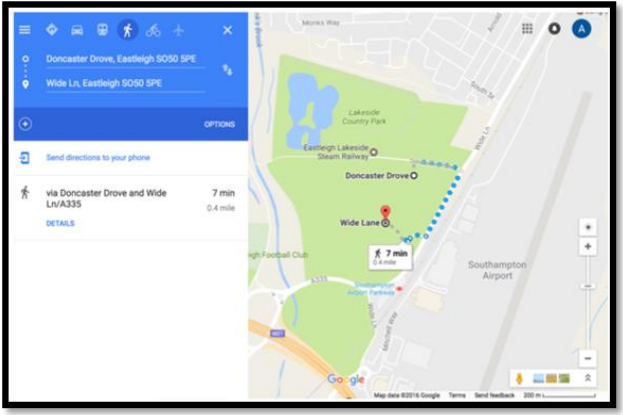

**Figure 2.** Google Maps suggested route for pedestrian between two points located in a park in Southampton

Figures 1 and 2 show the outputs of a routing query produced by two of the most commonly used routing services - OpenStreetMap (OSM) and Google Maps, respectively. The query was to find a route, in pedestrian mode, between point A and B, located in a grassland in Southampton, England. Both OpensStreetMap and Google Maps, project the origin and the destination points to the nearest street/road network and then calculate the path through using a graph-based algorithm. However, pedestrians and wheelchair users can both cross the grasslands and get from point A to destination B without traversing the network of roads. Wheelchair navigation, which may require additional topographic data, is not used by the native services/applications even if available.

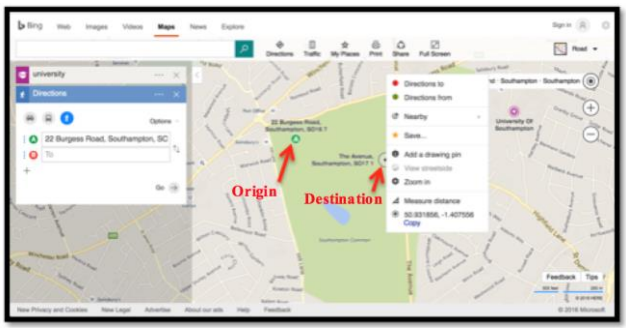

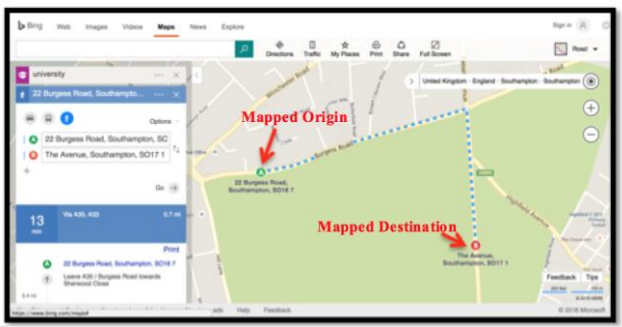

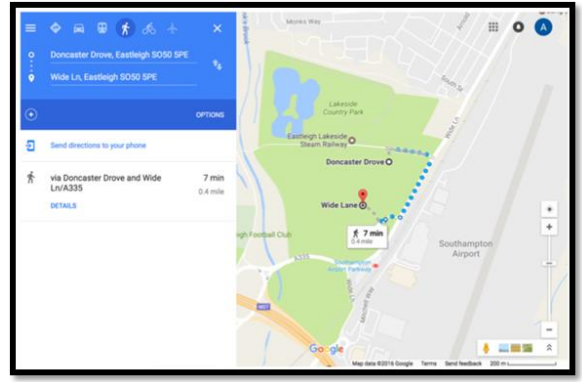

**Figure 3.** The user request (top) and Microsoft Bing Maps response (bottom), calculated based on the projected origin and destination points

Another widely used navigation service, Microsoft Bing Maps, does not even visualise the initial inputs of origin and destination points, if they are not located on the road network. Bing Maps projects the points onto the nearest road network and then visualises them, calculating the path for pedestrians based on the projected origin and destinations alongside the road network. Figure 3 shows the original origin and destination points given by the user (top figure) alongside the Bing Maps path between the projected origin and destination (Figure 3, bottom). For a user located off the road network in the public park (in this example), the Bing calculated route may be problematic as obstacles and barriers such as fences, may not allow them to access the road network as directed.

OSM provides wheelchair users with one of the most comprehensive databases, in terms of the attributes and "tags" required for wheelchair (and in general people with limited mobility) routing (Mobasheri, 2017, Zipf et al., 2016, Mobasheri et al., 2017a). However, most OSM routing services simply execute conventional graph-based path finding algorithms with the data in the OSM database (Mobasheri et al., 2017b), (Mooney and Minghini, 2017), as shown in Figure 1.

A crude solution open area routing could be to calculate a Euclidian straight line between the origin and destination, however it would fail to provide a usable route if there is an obstacle intersecting the straight line. Obstacles could include manmade features such as fences, walls and buildings requiring authorisation to cross, or natural barriers such as lakes and rivers. As Figure 4 shows, OSM suggests a path between the origin and destinations, which requires the user to cross Beach Lake and a railway line.

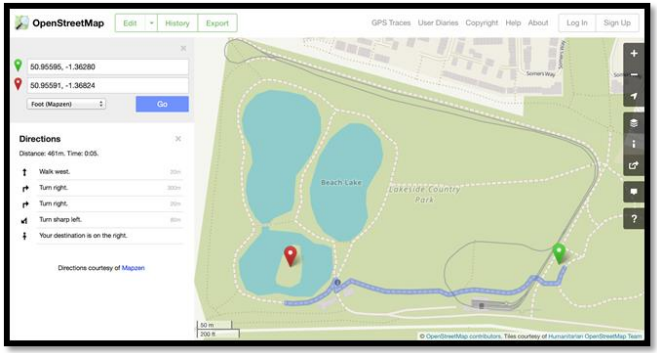

**Figure 4.** OSM suggested path requires the user to cross a lake and a bundle of train lines

These examples demonstrate that in open areas where there are no network constraints, some of the most widely used routing services could fail to provide a practical solution for pedestrians and wheelchair users, despite the existence of some algorithms for path finding in open areas (Hong and Murray, 2016a, Hong et al., 2016b, Giannelli et al., 2016, Donatelli et al., 2017).

There are a number of potential reasons why algorithms for path finding in open areas are not already employed by these routing services. One is that most of open area path-finding algorithms can only work with raster data (Goerzen et al., 2010), which is not fully compatible with the current vector/graph-based routing algorithms found in common multimodal navigation services. Choi et al., (2014) discuss the limitations of using raster-based path finding algorithms in a discrete vector network. A second reason is that the algorithms that work with vector data – mainly developed for robot and autonomous rovers – mostly focus on the obstacles to avoid rather than finding the optimum path. Visibility graph (Turner et al., 2001) – one such algorithm mainly designed for robots and Unmanned Aerial Vehicles (UAVs) can also be critiqued in that unlike humans these agents only have a single travel mode (Babel, 2014; Liu et al., 2013; Maekawa et al., 2010). For robots and UAVs the integration with network-based path finding algorithms and multimodality are not very important. In addition, the areas they operate in are assumed very complex and populated with obstacles to avoid. This is true for indoor environments and mid-air space where UAVs and drones move, and that is why several research projects have been completed to select and remove irrelevant obstacles from the scene to improve the performance of the path finding (Hong and Murray, 2013). As such, their performance and time to respond is arguably a big challenge for large areas. In addition, ignored by almost all of these algorithms are the impacts of the environment and personal preferences on the selection of paths by pedestrians and wheelchair users; weather, surface condition, and familiarity with the neighbourhood, could be important information in helping users decide to take a path.

This paper proposes and tests a novel path finding algorithm for open areas, which can be integrated with currently generated networks of roads and streets, for multimodal navigation purposes. The proposed algorithm creates an on-demand graph in the open area which can feature obstacles and barriers, and finds the most direct path between the origin and destination, avoiding obstacles and sympathetic to specialist needs such as those for wheelchair users. In order to calculate wheelchair-specific paths, the proposed algorithm weights the links based on distance and other important factors that wheelchair users usually take into account when they want to take a path. Such important factors are learnt from the patterns of movements extracted using trajectory mining and machine learning techniques.

The novelties and the contributions of the paper include: (a) proposing a finding algorithm in open areas for wheelchair users (applicable for pedestrians and bicycle users too) that can be integrated with currently used network-based algorithms to provide multimodal navigation service; (b) understanding the important factors for wheelchair users to take a path using machine learning and trajectory mining methods from the trajectories of movement of 98 participants in an experiments. The paper also evaluates and measures the compatibility of the proposed path finding algorithm's output with real-world choices of wheelchair users to assess the practicality and usability of the calculated paths.

This paper is structured as follows: Section 2 reviews the recent advancements and remaining challenges of wheelchair navigation and also the path finding algorithms. Section 3 will explain the proposed path finding algorithm and the process of trajectory mining and identification of the important factors in wheelchair routing is discussed. Section 4 evaluates the proposed algorithm outputs, both from users' perspectives and in terms of their similarity to the actual movements of wheelchair users.

## II. Wheelchair Navigation Services

Wheelchair users have more 'freedom' but a lower speed of movement than road traffic – akin to pedestrians in some ways. The ability to move across open spaces makes them less constrained than road traffic but more vulnerable than pedestrians to environmental conditions. For example, the topography and weather can have an important impact on wheelchair users' choice of routes. In addition, any changes in the pre-planned path can impose a significant amount of energy and time to be spent. Ferrari et al (2014) showed that half of the trips in London become 50% longer for people with mobility impairments as they need to take longer trips or change the planned routes to avoid obstacles and inaccessible areas.

Open areas, where the wheelchair users are moving, are not as well-regulated as the roads and streets, in terms of activities that can be carried out in such areas, and this makes the environments more open to temporary changes and obstacles compromising the reliability of the planned routes (Neis, 2015). The importance and particular requirements of wheelchair navigation has made it a topic of research even before navigation applications built into smartphones became popular amongst pedestrians (Levine et al., 1991, Simpson et al., 1998, Simpson et al., 2002, Harrison et al., 2004, Kasemsuppakorn et al., 2009, Kasemsuppakorn and Karimi, 2009). This is mainly due to the need for assistive services to help the wheelchair users in their everyday mobility activities. For almost the same motivations, this paper focuses on wheelchair navigation services.

### A. Related Works on Wheelchair Users Demands

With the development and popularity of smart phones and mobile computing, several applications ('apps') and services have been developed and many research projects have been working on providing the wheelchair users with better navigation experiences (Karimi et al., 2014a; Dalsaniya et al., 2016; Salhi et al., 2016; Krieg-Brückner et al., 2013; Harriehausen-Muhlabauter, 2014; Oliveira et al., 2016; Cardoso et al., 2016). Different aspects of wheelchair navigation services, including positioning and localisation (Salimi et al., 2013) and (Bejuri et al., 2013) have been studied, facilitated by the ability to mount additional sensors to the wheelchair (Alasfour, 2015).

In addition to the location of the user, accessibility maps are one of the most important input data for wheelchair routing and navigation. Beale et al., (2013), Qin et al., (2016), Ponsard et al., (2016), and Karimi et al., (2014b) all focused on the topic of accessibility maps, which provide wheelchair users with obstacle data and 'impossible to take' routes. The quality of accessibility maps can have a significant impact on the quality of the final service. Kim, (2016) proposed and implemented a system, which can detect obstacles using a combination of a vision and eight ultrasonic sensors, producing occupancy grid, maps. Although this system can map obstacles on the fly, the output map is a raster (grid) while many path-finding algorithms are vector/graph-based and conversion is a challenge. Also, the on-the-fly mapping requires several sensors, which could be financially and computationally expensive to implement. It also cannot detect ramps that are too steep, poor surfaces and the absence of curb drops– all of which are relatively important factors for wheelchair routing.

Some researchers and service developers focused on routing algorithms for wheelchair users. Tajgardoon and Karimi (2015) simulated the accessibility of a route/segment for way-finding purposes. Although this simulation considers both obstacles (such as stairs and fences), and important facilities (such as ramps or curb drops) to label a segment as uncomfortable or inaccessible (such as the absence of or ramps that are too steep, poor segment surface, no curb drops or blocked curb drops), it is only limited to pavements in urban environments and paths for open areas are not resolvable. It also visualises the results of simulations through heat-maps, which are rendered as raster grids and so are a challenging format to be fed into currently available network-based routing algorithms for multimodal navigation. In addition, the preferences and priorities and their corresponding values are user generated inputs, which could be affected by semantics and the interpretation of individuals. Kasemsuppakorn et al. (2015) proposed a personalised routing service for wheelchair users based on user preferences. The calculated paths are, on average, almost 15%, longer than the shortest path but with slope and surface conditions more friendly to wheelchair users. Hashemi and Karimi (2016) proposed and implemented a collaborative approach for personalised wheelchair way-finding, which uses conventional routing services (and so conventional path finding algorithms) but receives updates regarding personal preferences on the network segments. This paper uses machine learning methods and trajectory mining techniques to infer such important factors (preferences) and their priority values based on real movement data from wheelchair users.

### III. Open Area Path Finding Algorithm

To calculate the path between origin and destination points, usually a shortest (or in general the lowest 'cost') path algorithm is used. The shortest path algorithms for pedestrian, wheelchair and car navigation usually work with vector or graph data (Hong and Murray, 2013) while the raster-based algorithms (Van Bemmelen et al., 1993, Antikainen, 2013, Chang et al., 2003, Baek and Choi, 2017) are mainly applied for robot and UAV navigation and game applications.

Choi et al. (2014) proposed and implemented a path finding algorithm providing the least-cost path on a continuous surface as well as a discrete vector network. As most of routing services uses graph-based algorithms, raster-based path finding algorithms may not provide the interoperability and support the multimodality. In addition to the need for multimodality, the movement cost across large areas could be unchanging meaning raster based routing algorithms may not calculate one single recognisable shortest path.

For a connected graph, multiple paths can exist between any pair of nodes. Often, we are interested in the path that has the shortest path or minimum cost. Many shortest path algorithms, such as Dijkstra's algorithm, the Bellman–Ford algorithm, the A* search algorithm, the Floyd–Warshall algorithm, Johnson's algorithm, sparse graphs and Perturbation theory (Li et al., 2004, Deo et al., 1984, Fu et al., 2006, Jagadeesh et al., 2002) are graph-based. These graph-based algorithms are perfectly compatible with transportation networks, where the segments of streets, metro lines, motorways, etc. can be considered as links and junctions or stations can be mapped into a set of nodes. However, the projection of an open area, as a continuous field, into sets of nodes and links may not seem as natural. Hong and Murray (2016a) and Goerzen et al. (2010) studied open area path finding approaches, such as visibility graph (Turner et al., 2001, Wang et al., 2017) and mesh navigation (Niederberger et al., 2004), and compared the representational issues, optimality, precision, computational requirements and resulting path efficiency.

Most of these approaches have been designed for robots or agents to avoid obstacles, in complex environments such as indoor spaces. They are particularly successful in robotics and game applications, however, the emphasis of many of them (e.g. visibility graph algorithms) is to avoid obstacles, rather than to search for traversable corridors/polygons. For wheelchair navigation, traversability is the major emphasis.

Many other algorithms (e.g. navigation mesh) do consider walkability or traversability but simply as an additional piece of information attached to the data and so treat traversability in the same way as length or impedance. This paper proposes an algorithm, which considers both traversable areas and obstacles at the same level of abstraction. Considering both obstacles and traversable areas at the same time could potentially add to the complexity of the algorithm if the number of the obstacles and/or the open area size is large. To minimise such potential complexity, a hierarchical approach is proposed that takes (an approximation of) the obstacles into account only when needed.

Hahmann et al. (2017) compared several open area path finding algorithms, including Grid, Spider-Grid, Delaunay, Voronoi, Skeleton, and Visibility from four perspectives of route computation time, routing quality, number of additional created graph edges, and additional route computation time in three study areas. Graser (2016) also compared four path finding algorithms: Medial Axis, Straight Skelton, Grid and Visibility in open areas. Both studies agreed that the Visibility graph algorithm provides a more realistic graph with fewer additional edges.

For any arbitrary origin and destination points inside a polygon, which are represented by s and t respectively in this paper, the algorithm should be able to calculate a path avoiding obstacles and holes. The main steps of the process of generating a graph from an open area (i.e. from a continuous space) for path finding purposes are shown in the flowchart in Figure 6.

First, the algorithm takes the polygon, origin, and destination and then creates a graph with the two main elements of nodes and links, N (nodes) and L (links) respectively. Suppose $G = (N, L)$ denotes a graph to be created for the polygon in the plane having $h$ holes (obstacles) and a total of $m$ vertices (generating the boundary of the polygon). $N$ is a set of vertices, $N = \{n1, n2 \ldots nm\}$, including vertices of the boundary of the polygon, which includes the vertices of holes inside the polygon if there is one and/or the geometry is not a simple polygon (see Figure 8 as an example). $L$ is a set of links/edges, including segments of the polygon boundary, which also includes the segments/edges of the holes. Given a start point or the origin, s∈G, and an end point, the destination, t∈G, the problem is to find the shortest path inside the polygon from s to t.

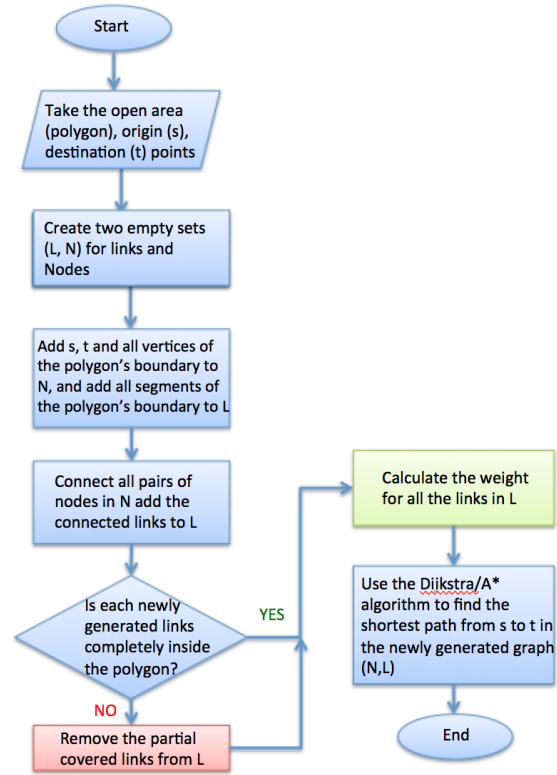

**Figure 6.** The graph creation and open area path finding algorithm flowchart

Figure 7, left, illustrates an open area polygon in which the path between the start (s) and the destination (t) points is requested. The polygon has one obstacle, and an entrance as two common constraints of open areas. The direct Euclidean line could not be taken due to both the polygon geometry and the existence of the obstacle.

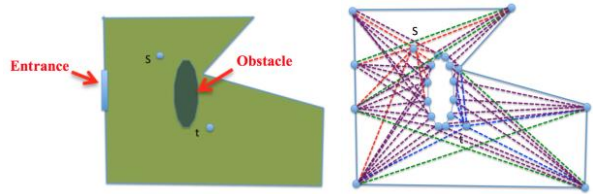

**Figure 7.** The open area polygon, where the shortest path is requested between the indicated origin and destination points (left), and the generated graph using the boundary points and constraints (right)

To generate the graph, every pair of vertices comprising the polygon boundary and the origin and destination points are connected and the connecting lines are stored in a temporary set, called "temporary edges". The connecting lines, which are partially or completely outside the polygon, are excluded from the set. For example, in Figure 8 the direct line between the origin and destination crosses the boundary of the polygon and so the topological relationship between the connecting line and the polygon will not be "completely inside". The remaining links in the set of "temporary edges",

2 VOLUME XX, 2017

i.e. the links that are not crossing the boundary of the polygon, are removed.

If a polygon includes some obstacles, then the vertices and nodes of the obstacles are also added to the initial sets of N and L. And so, their boundaries and vertices are included in the process of generating the graph. The storage of a hole/obstacle in different databases varies (e.g. vertices of the boundaries of holes in a counter-clockwise order). Figure 8 shows the results of the shortest path for the graph created in Figure 7.

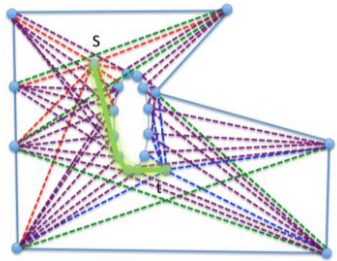

**Figure 8.** The shortest path calculated using the algorithm illustrated in figure 7 is highlighted in green

If open area polygons representing objects like bounded urban parks have constraints such as entrances or exit gates, then where either the origin or destination point is located outside the open area, it is very likely that the entrance/exit is linked to a network of roads, pedestrian pathways, or streets so there is no need to generate a new graph to join the internal open area graph. Where origin or destination points are located outside an open area, then simply the best path between the origin/destination and the centre of the entrance/exit gates is calculated and the centres of the entrance and/or exit gates are added to the set of nodes, as a new origin or destination. The best path between the point(s) outside the polygon and the centre(s) of the gate(s) will be joined and the best path calculated inside the polygon using the proposed algorithm. In the unlikely case of having no connection between the open area and the street network, there will be another stage in the initial flowchart shown in Figure 6. This stage generates another graph between the out-of-polygon points and the centre point of the entrance/exit gates. The two graphs are linked to each other through this single node.

This algorithm assumes open areas are single polygons that could have holes. However, for multi-polygons, if both the origin and destination points are not located in one single element polygon, then to be able to solve the path finding problem by the proposed algorithm, it can be reasonably assumed that there are some links (e.g. pathways) between the open area polygons. These links are also taken into account when generating the graph.

While this approach considers both obstacles and open areas the computational expense it adds to the system, limits its applications to relatively simple scenarios e.g. where the open area boundary does not consist of too many vertices and so the generated graph is not too dense. However, given the proposed application for wheelchair trips which will not cover large distances, such scenarios may either only happen in rare cases or be solved through the generalisation of some of the vector data. While this algorithm can calculate a path for small areas and/or where the shapes are simple, we also propose a hierarchical process to generate the graph for only a sub-area and include the obstacles only when and where it is needed. The proposed algorithm can perform better in larger and/or more complex areas with potentially more obstacles as not does not include all vertices for every single step. Unlike the previous algorithm where an obstacle can be simply considered as holes in a polygon and add some vertices to the node set, the proposed (hierarchical) approach considers obstacles as polygons themselves.

It starts with direct line between the origin and the destination and if this line overlaps with the holes/obstacles or the outside of the study polygon then try to consider the vertices of the obstacles that overlap with the path. This continues until a polyline is found that does not cross any of the obstacles. This process in high level of abstraction is shown in Figure 9.

This algorithm firstly takes the origin and destination points as two nodes and so the connecting line between these two nodes is stored temporarily as a link. While the link is completely located inside the open area, there could be a blockage by the obstacles. If the link crosses Minimum Bounding Rectangles (MBRs) of any obstacle, then the topological relationship between the link and the obstacle polygon is checked. If the overlap just exists between the MBR(s) and the link, and not with the obstacle polygon(s), then the link is selected as the shortest path. However, in more complicated scenarios, the link can cross the obstacle(s), and in such cases, the MBR(s) corners are added to the set of nodes and the set of links is updated according to the same rule of full inclusion in the open area polygon, which have been explained before.

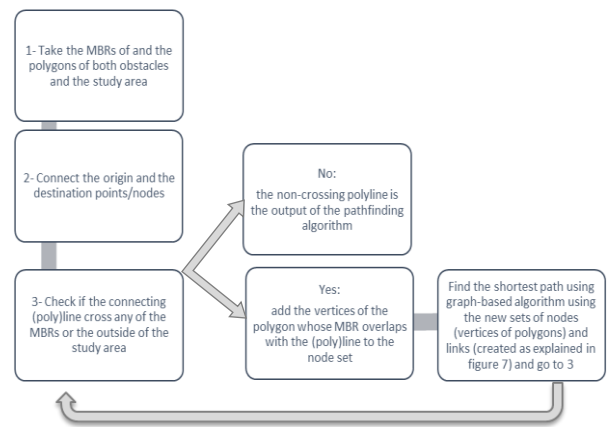

**Figure 9.** The (high level of abstraction) hierarchical open area path finding algorithm

The path finding algorithm minimising the cost is again applied, this could follow the same algorithm used in the previously explained approach, i.e. weighting the links based on the important factors for routing and selecting the lowest cost path using an algorithm such as A*. The selected path is again checked to find any overlapping MBR(s) and potential obstacle polygons. This continues and at each iteration, the sets of nodes and links are updated and populated by the nodes and links identified by the overlapping obstacles identified in the previous iteration. The links are generated following the same rule of full inclusion, i.e. the connecting lines of every pair of nodes are stored as links, except the links which are not completely inside the open area polygon. This continues until a path is found with no overlap with the obstacles located in the open area. In the worst-case scenario, the path is identified at the last iteration, where all the MBRs are included. In this case the first direct link does not qualify as a link, and the vertices of the polygon, in addition to the origin and destination points, are considered as nodes and the graph is generated accordingly. This hierarchical algorithm is shown in the Figure 10.

This hierarchical process will add the four points of the MBR corners, instead of all the vertices of each obstacle boundary. Therefore, the complexity and density of the generated network is much lower than the network generated by the previous algorithm. This becomes very important where there are more than just a few obstacles to avoid and/or the obstacle boundaries are stored with many vertices. Predictably due to the approximation of the obstacles with their MBRs, the results of the two algorithms do not necessarily match entirely. Hong and Murray (2013), proved that using a convex hull guarantees the shortest path, and the generated graph here in both scenarios is based on utilising convex hulls but does not consider obstacles as they are and uses the MBRs to approximate and improve the efficiency and performance.

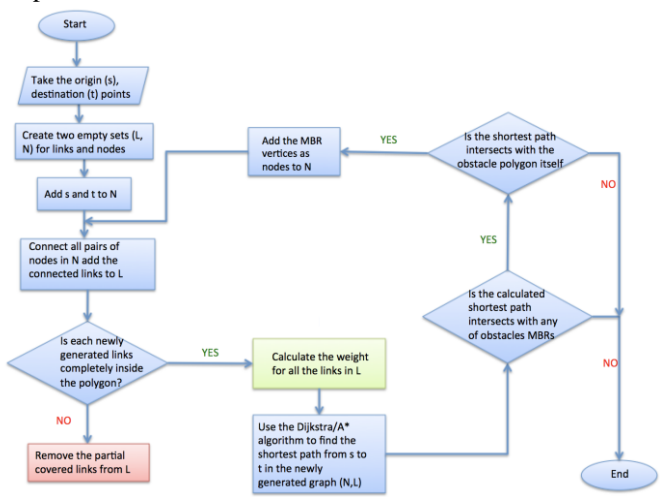

**Figure 10.** Flowchart for open area path finding using step-wise obstacle exclusion

## A. Weighting Links

For wheelchair navigation, in addition to distance, there are other important parameters, such as the surface condition, slope and width of the passage, which need to be considered/optimised. In order to identify the important parameters, several models for predicting a correction value based on the environmental and spatial factors can be explored. To exemplify we utilise a dataset of the trajectories of volunteer wheelchair participants captured by a tracking smartphone application. The data capture was conducted during May 2016 – Oct 2016 and the dataset includes 286 whole-day trajectories and 371 "open area only" trajectories. An "open area only" trajectory is defined as a movement trace that either partially or completely overlaps with an open area, i.e. a polygon with no networks of roads or passages, and a change in course is no longer limited to observed road or pavement areas. After the data capture phase, the participants scored their route choices from the best (5) to worst (1). Some alignment of perceptions was facilitated through discussions with the participants and a final normalised score assigned to route segments as a combination of participants personal experiences and a collective evaluation of each.

Captured data, trajectories and the associated scores, are randomly divided into two sets for two separate purposes (a) recognising the important factors in determining a good route and (b) service evaluation. The dataset which is used for important factor identification is itself randomly divided into two sets - a training and a control dataset.

After data collection and data preparation, training data (that include the trajectories, and the participants and movement data including the age, gender, time of day (hour), day of week, length of the journey, total length of journeys in a day (sum of length of all journeys during the same day) and weather condition (cloudy, rainy, etc.), slope, width, surface condition, are used to identify the important features. The variables such as age, gender, time and length are considered as independent variables (features) and the spatial and temporal proximity to the participants actual movements are considered as the target variables.

The speed of movement around the whole network can be estimated using the training data and a range of estimation models. Various algorithms with parameter optimisation are tested and cross-validated using the control dataset. Table 1 shows the prediction accuracy of the different algorithms using R2 (coefficient of determination) along with the training time in milliseconds. The model training and validation are done on a desktop machine with 16GB RAM, Core i-7 4600U quad core CPU, and 2.70 GH clock speed.

TABLE I
TRAINING ACCURACY AND DURATION OF VARIOUS PREDICTION MODELS. DURATION TIME IS IN MILLISECONDS AND IS RECORDED FOR 10-FOLD CROSS VALIDATION.

| Prediction Method | Coefficient of determination (R2) | Training Time (milliseconds) |
|---|---|---|
| OLS (Ordinary Least Squares) | 0.5441 | 102 |
| LASSO | 0.8612 | 421 |
| **LARS** | 0.8567 | 456 |
| **Ridge Regression** | 0.8618 | 432 |
| **SVM (Support Vector Machine) with linear kernel** | 0.8212 | 691 |
| **SVM with Polynomial kernel** | 0.8061 | 819 |
| **SVM with RBF kernel** | 0.8659 | 1001 |
| **Random Forest** | 0.9167 | 4702 |
| **Gradient Boosting Decision Tree** | 0.9181 | 4328 |

As illustrated in Table 1, the ensemble models, random forest and gradient boosting decision tree are most accurate but with a corresponding time cost. The penalised linear regression models such as Least Angle Regression (LARS), Least Absolute Selection and Shrinkage Operator (LASSO) and ridge regression are relatively accurate while being faster in the training phase and offer a good compromise in comparison with the other slightly more accurate and more complex models. In general, penalised linear models are simple and often provide an adequate and interpretable description of how the input features affect the output. For prediction purposes, they can sometimes outperform nonlinear models, especially in situations with small numbers of training cases, low signal-to-noise ratio or sparse data (Hastie et al. 2009).

The importance of training duration for this application is twofold: firstly, in the final version of the route finding mobile 'app', the regression model will be retrained continuously to provide each user with personalised route while updating optimum routes from newly captured and stored trajectory data. The app keeps recording the data and these are used for tuning (modifying) the parameters of the initially deployed model. Given this, the speed of model training is important and penalised regression models are the fastest models for training. Secondly, more complex models (like ensemble models and SVM) need more processing power and time for retraining and where this would be carried out on a mobile device, a very real issue of battery life emerges. More complex models use more battery life than the simpler models.

The important features and conditions which encourage or discourage wheelchair users to take a path include its surface condition (man-made material/vegetation, elasticity, unevenness degree etc.), , minimum width, maximum slope, the time of the day and the associated lighting condition along the route, the weather. These interact with other variables such as the length of the journey and the age or gender of the participant.

Significantly, the coefficients of the least important features (age, and gender) are almost zero. This is important since the prediction model in the final service can be used without profile information, e.g. age and gender. The normalised importance value of the top five important features and the couple of least important features are shown in Figure 11.

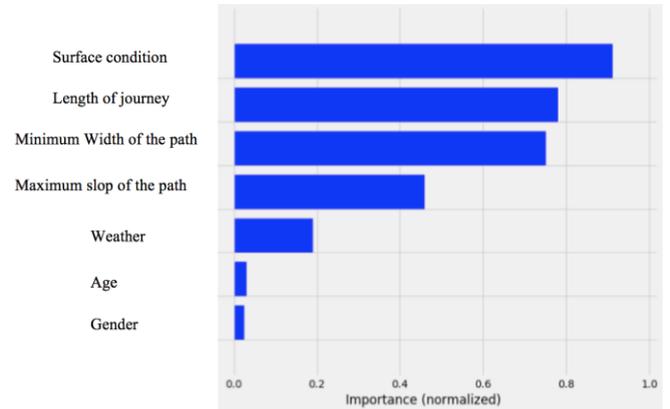

**Figure 11.** The importance of features (attributes) calculated by the prediction model for wheelchair paths.

Based on the identified features, it is now possible to weight the links of the generated graph. The weight of each link is a function of five parameters: the distance between the start and end node of the link (length of the edge), the maximum slope of the link, the narrowest width of the link, the poorest surface condition rate, and the weather condition. The weight of each link is a weighted average of these feature values for/at each link. As explained before, the values of each of the important features are assumed constant for each link and if there is any change (for example the surface condition or the width value), then the link must split into atomic (single valued) links and one (or more) node(s) at the break will be added to the set of nodes and the graph will be re-generated. However, it may also simplify the case and calculate the weight based on the minimum value of the feature, i.e. the poorest conditions. For example, the highest slope, the narrowest width, the poorest surface condition rate, are considered for weighting.

Figure 12 shows an example of the output of the calculated paths between the same origin and destination points by Google Maps (top left), OSM (top middle) and BingMaps (top right) and the proposed algorithm (bottom left) along with the actual traced path by the user (bottom right). The output of the calculated paths between an arbitrary origin and destination points is much closer to the route that the participants who are familiar with the area would take if there was no navigation service available. This may indicate the path finding service is closer to the way human would find their ways, and of course a huge improvement to the currently available routing services provided by Google Maps, OSM and BingMaps. Details of the qualitative and quantitative evaluation of the path finding algorithm is presented in the next section.

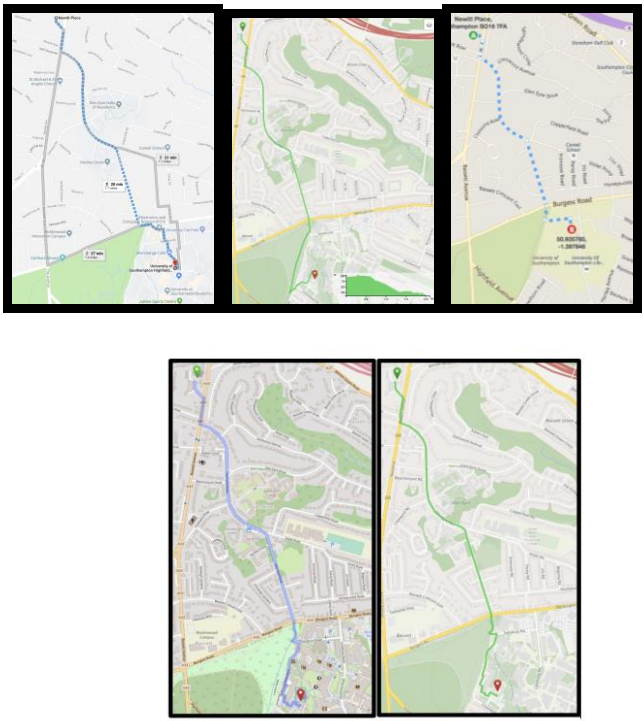

**Figure 12.** The output of routing services of Google Maps, OSM, Bing Maps, the proposed open area path finding algorithm and the actual traces of the user (from top left to bottom right)

## IV. Evaluation

In order to evaluate the accuracy and usability of the proposed path finding algorithm, an experiment was conducted. This experiment allows us to evaluate the results of the algorithm both qualitatively and quantitatively as both traces of movement and also the participants comments on the suggested path are analysed.

19 participants - 11 female and 8 male who are aged between 18 to 49 - were recruited and they were provided with the paths calculated by the proposed new algorithm built into a smartphone application. The participants included 13 wheelchair users (excluding motorised and powered wheelchairs), 2 people who need walkers or other mechanical tools for their daily mobility, and 4 people who carried (infant) pushchairs. Their movement trajectories were captured by a location detecting element of the application ('app') for 15 days; for 7 of these they followed the paths suggested by the new algorithm. The participants had the option to pause the sensing module but to finish the study a set number of tracking hours needed to be accomplished. Participants had smartphones running both Android (8 participants) or iOS (11 participants) operating systems and the tracking app was available for both groups with the same interface. Figure 14 shows the tracking app interface running on an Android device. The app evaluated the satisfaction level (experience sampling) of the users and compared their movement trajectories when they were not using the route-finding app with the output of the proposed algorithm along with the output of conventional network-based path finding algorithms.

The app stores the position of the participant every two seconds and the raw data are sent to the server as soon as an internet connection is available. Using a developed ArcGIS Add-in, explained in detail in (Basiri et al, 2016b), raw data are pre-processed, error/noise removed, positional data are compressed and simplified, and the final trajectories are generated and visualised. In the pre-processing step, raw trajectory data are not being matched against other map data in order to minimise any data exclusion which could occur with movement in open areas. This can potentially degrade the reliability of the trajectory as non-matching segments may include vertices with high location uncertainty. However, this has been addressed by giving an edit option to the user (see Figure 13), where they are able to redraw their trajectories if they are not accurately captured automatically. In addition, the relatively high speed of the location sampling rate (2 seconds) with respect to the pedestrian and wheelchair users speed of movement helps to minimise error and noise detection in the later process, using an enhanced Douglas-Peuker algorithm (Basiri et al., 2016b). The sampling rate is intentionally selected to be high, assuming that the relatively slow movement of a wheelchair user cannot change the geometry of the trajectory dramatically in just 2 seconds. Even if the sensed position data include some error, the noise filtering step should help correct it.

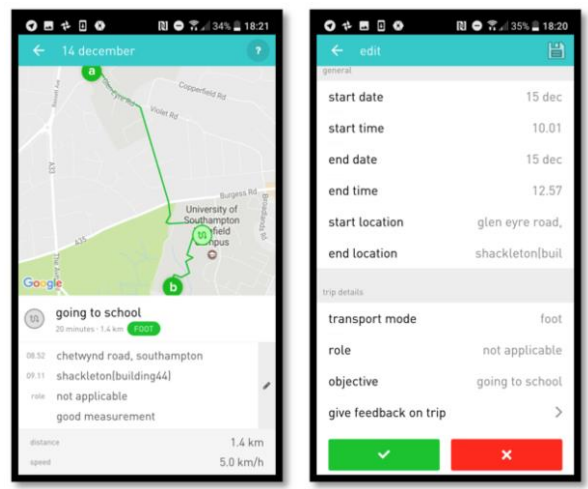

**Figure 13.** The tracking app and the edit option for the participant to modify the geometry and attributes of the route

In addition to the (sequences of) location and time, weather, surface condition, the purpose of the trips, duration and date, and other feedback of the participants are captured or inferred by the app. These values are used for training and tuning of the models. These values either come from Application Programming Interfaces (APIs) such as a public weather API or assigned by the app using pre-set if-then rules that match the values, e.g. surface condition, against the location and time. For example, using a simple map matching the app can assign an initial value to the surface condition, e.g. muddy grassland or asphalt, based on current location/feature user is at. However, the users have the option to edit both the geometry of the pre-processed trajectories and the attributes, such as weather condition or date and time for each trip that is assigned to each trajectory. The captured data during this experiment are used to extract the important features and also will be used as a control set for the accuracy estimation of the results.

In addition to the 7 days of data capture, for another 8 consecutive days the implemented path finding algorithm provides navigational instructions to the participants. After 15 days of app usage, the users' feedback shows that a large majority of the participants (16 participants out 19), were either "extremely satisfied" or "satisfied" with the provided paths. 14 confirmed that the suggested paths are "identical" or "very close" or "similar enough to be replaceable" to the paths they would usually take.

Captured movement trajectories were compared with two other sets of routes: firstly the output routes calculated by currently available routing services; and secondly the movement trajectories of the very same participants during the second week of the study when they were not following routes suggested by the routing algorithm.

The comparisons of unguided movements with either of the suggested sets is carried out in relation to two dimensions - user satisfaction and the distance between the trajectories. For satisfaction, this means that where the users' experiences are sampled during the experiment both with, and without path suggestion, the level of satisfaction is considered as a weight, and a score between 0 and 5 recorded for each trajectory segment. For trajectories, the similarity of any two routes corresponding to the same origin and destination is calculated using three different measures: (a) Closest-Pair Distance (CPD), (b) Longest Common Sub-Sequence (LCSS), and Trajectory-Hausdorff Distance ($DHaus$).

Closest-Pair Distance uses the minimal distance between the points in two trajectories ($A$, $B$), with sequences of points of $p$ and $p'$, to represent the similarity of trajectories, (see Equation 1). This similarity measure is relatively simple to calculate, which allows the comparison of the trajectories with very high number of vertices. In this experiment the sampling rate has been set at up to two seconds and so due to the nature of open areas, where there are no roads and passages, the map matching cannot filter many of them. Although the trajectories are simplified using the enhanced Douglas-Peucker algorithm (Basiri et al., 2016b), the intention is to keep as many points as we can to make sure the movements of the participants are monitored precisely. As such, having so many vertices require a simple similarity measure, and CPD serves this purpose.

$$CP(A, B) = min p \in A, p' \in B D(p, p') \qquad (1)$$

To deal with noise in the datasets which can make two trajectories either unrealistically close or far from each other, another measure is also used to have a better understanding of the similarity of the trajectories. The LCSS-based Distance measure allows some noise points to be skipped while calculating the distance of the trajectories, using a threshold, $\varepsilon$, to control how far we can go in order to match one point from a trajectory to a point in another trajectory.

The third similarity measure, Trajectory-Hausdorff Distance ($DHaus$), is used to measure how similar the open area suggested path is to the actual movement of wheelchair users. It can also be used for comparing the similarity of the currently available routing service paths and the proposed algorithm outputs. It is a weighted summation of three components: (a) the aggregate perpendicular distance ($d\perp$) that measures the separation between two trajectories; (b) the aggregate parallel distance ($d//$) that captures the difference in length between two trajectories; and (c) the angular distance ($d\theta$) that reflects the orientation difference between two trajectories (Lee et al., 2007), see Figure 14.

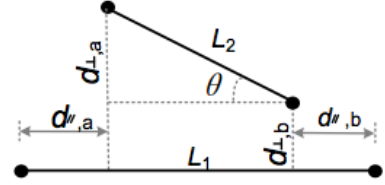

**Figure 15.** Trajectory-Hausdorff Distance ($DHaus$) components

This study only considers the average of $DHaus$ calculated for the first and the last segments of the trajectories. This means $DHaus$ is calculated for the segments sharing either a start or end point and is driven by the fact that trajectories can have different number of vertices and segment lengths for the same origin and destination. Considering the average of $DHaus$ for the first and last segments also means that the perpendicular distance is excluded (i.e. becomes zero) as there is a sharing vertex in the comparing segments, joining the two segments. $DHaus$ can be formulated:

$$DHaus = w1\, d// + w2\, d\theta \qquad (2)$$

Where $d//$ is the distance of the non-sharing vertices, and $d\theta$ is the projected distance of one of the comparing segments on the other comparing segments (i.e. $d\theta = \|L2\| \cdot sin\theta$ – see Figure 15), and $w1$, $w2$ are the weights which were equally considered as 1.

The results of the comparisons show the success of the proposed algorithm, both in terms of participant satisfaction and also in terms of the similarity of the trajectories suggested by the algorithm with respect to actual wheelchair user movements. 17 participants, out of total 19, ranked the suggested paths as "good enough to be taken on daily basis" and 14 of them confirmed that the suggested path is replaceable by the paths they would take if there had not been any recommendations.

The similarity measures, of Closest-Pair Distance Longest Common Sub-Sequence, and Trajectory-Hausdorff Distance, show a significant improvement in the path finding as the proposed algorithm output was much closer to the actual movements of participants, in comparison with the currently available routing services of OpenStreetMaps and Google Maps. CPD, LCSS, and $DHaus$ measures show that the open area path finding algorithm can produce 76.4%, 80.3%, and 89.8% (respectively) geometrically closer results to the user's trajectories in comparison with the suggested path by Google Maps and 78.5%, 82.5% and 92.6% closer than OpenStreetMap's. This has been demonstrated in Figure 12.

## V. Conclusion

This paper proposes and evaluates a novel open area path finding algorithm for wheelchair users, which considers their particular requirements and demands, identified based on trajectory mining techniques. The proposed algorithm generates a graph over the open area based on a hierarchical approach considering both obstacles and traversable areas. This graph can be integrated with currently used graph-based algorithms to provide multimodal navigation services. The graph is weighted based on important features for wheelchair users, such as slope, surface condition, weather, and width. These parameters are recognised using machine learning and trajectory mining techniques. The output of the algorithm was compared with the actual movements of wheelchair users to see how close the calculated path could approximate their movements. The open area path finding algorithm has also been compared with the outputs of currently available routing services using a suite of similarity measures (CPD, LCSS, and $DHaus$) and this demonstrated that the proposed algorithm produces routes that are closer to those that would normally be chosen by wheelchair users than routes suggested by commonly used applications such as Open Street Map and Google Maps. In addition, during an experiment, the majority of the participants confirmed that the output of the open area algorithm was acceptable enough to replace the path they would take if there were no instructions.

In the future the proposed algorithm could be customised for other 'active' travel modes, such as bicycle and pedestrians as they may have other important preference which can be learnt by the machine learning techniques (quiet streets, avoiding busy junctions etc.). In addition, the integration of the wheelchair mode with the other travel modes and the routing services optimised for them need to be tested (in terms of performance and scalability) in future.